\documentclass[11pt]{article}
\usepackage{teschool,epsfig,amssymb}
\usepackage{graphicx}
\usepackage[dvips,usenames]{color}

\def\Journal#1#2#3#4{{#1} {\bf #2}, #3 (#4)}

\def\PLB{{\em Phys. Lett.}  B}
\def\PRL{\em Phys. Rev. Lett.}
\def\PRD{{\em Phys. Rev.} D}

\def\JPG{{\em J. Phys.} G}
\def\NPA{{\em Nucl. Phys.} A}
\def\PRC{{\em Phys. Rev.} C}
\def\EPJC{{\em Eur. Phys. J.} C}

\def\be{\begin{equation}}
\def\ee{\end{equation}}
\def\bea{\begin{eqnarray}}
\def\eea{\end{eqnarray}}

\def\beqstar{\begin{eqnarray*}} \def\eeqstar{\end{eqnarray*}}
\def\bal{\begin{align}}
\def\eal{\end{align}}
\newcommand{\p}[1]{(\ref{#1})}

\newcommand{\tit}{\textit}

\begin{document}
\vspace*{4cm}
\title{PHYSICS OF THE ALICE EXPERIMENT AT THE LHC}

\author{ A.A. ISAYEV }

\address{Kharkov Institute of
Physics and Technology, Academicheskaya Str. 1, Kharkov, 61108,
Ukraine\\Kharkov National University, Svobody Sq., 4, Kharkov,
61077, Ukraine}

\maketitle\abstracts{Some physical aspects of the ALICE experiment
at the LHC are considered with emphasis on possible probes of
quark-gluon plasma, created in ultrarelativistic heavy ion
collisions.}

\section{Introduction  }
%\subsection{Producing the Hard Copy}\label{subsec:prod}

A Large Ion Collider Experiment (ALICE)~\cite{JPG1,JPG2} is the
dedicated heavy ion experiment at the CERN Large Hadron Collider
(LHC). The nucleon-nucleon  center-of-mass energy for collisions
of the heaviest ions at the LHC ($\sqrt s = 5.5\,\mbox{TeV}$) will
exceed that available at Relativistic Heavy Ion Collider (RHIC) in
Brookhaven by a factor of about 30, opening up a new physics
domain.
  ALICE will investigate equilibrium as well as
non-equilibrium properties of strongly interacting matter in the
energy density regime $\varepsilon\simeq 1-1000\,\mbox{GeV fm}^{
-3}$. At such huge energy densities a new state of matter,
quark-gluon plasma (QGP), consisting of deconfined quarks and
gluons, is expected to occur.  ALICE is also aimed to investigate
 proton-proton, proton-nucleus collisions as well as collisions
of lighter ions in order to get an important benchmark for
collisions of heavy nuclei and to separate phenomena truly
indicative of the hot and dense matter from other contributions.
The focus of heavy ion physics is to study how collective
phenomena in nuclear matter under conditions of extreme density
and temperature  emerge from the microscopic laws of elementary
particle physics.

The quantum field theory of the strong nuclear force is Quantum
Chromo-Dynamics (QCD). The color charge of quarks comes in three
variants: red, green and blue. In particular, the gauge bosons -
gluons,  carry themselves a nonzero color charge, giving rise to
the gluon self-interactions. The QCD Lagrangian reads \be {\cal
L}=\bar q(i\gamma^\mu D_\mu-\hat
m_q)q-\frac{1}{4}F_{\mu\nu}F^{\mu\nu},\label{lagr}\ee where $q$
and $\bar q$ denote the elementary matter fields, quarks and
antiquarks. All flavor, color and spin indices in Eq.~\p{lagr} are
omitted. The mass matrix $\hat m_q = {\rm
diag}(m_u,m_d,m_s,m_c,m_b,m_t)$ is the  diagonal matrix in the
flavor space, containing  quark masses. The interaction of the
quarks is encoded in the covariant derivative,
$D_\mu=\partial_\mu-igA_\mu$, where $A_\mu$ denotes the gluon
field,  and the gauge coupling $g$ quantifies the interaction
strength. The gluons are massless particles with spin 1,  carrying
color charge in 8 different charge-anticharge combinations, e.g.,
red-antigreen, red-antiblue, etc. The commutator term in the gluon
field tensor $F_{\mu\nu}=\partial_\mu A_\nu-\partial_\nu
A_\mu+g[A_\mu,A_\nu]$ describes the gluon self-interactions.

The gluon self-interactions lead to a remarkable property of QCD,
namely, the "anti-screening" of the color charge in the vacuum:
the virtual quark-gluon cloud around a color charge in the
polarized vacuum  induces an increase of the effective charge with
increasing distance. This property is closely related to the
running coupling constant  of QCD, \be
\alpha_s(Q)=\frac{12\pi}{(33-2n_f)\ln\frac{Q^2}{\Lambda^2}},\label{alpha}\ee
$n_f$ being the number of active quark flavors at the given
momentum transfer $Q$ (the number of flavors with $m_q\leq Q$),
$\Lambda_{QCD} = 0.2\, {\rm GeV}$ being the regularization
constant of QCD. Eq.~\p{alpha} illustrates the property of
asymptotic freedom: $\alpha_s\rightarrow0$ as $Q\rightarrow\infty$
meaning that  at small distances $r\sim 1/Q$ the quark and gluon
interactions become weak and, hence, the perturbation theory on
$\alpha_s$ can be developed. Oppositely, at scales $Q\sim
\Lambda$, QCD becomes strongly coupled and perturbation theory
cannot be used. In the strong coupling regime, new nonperturbative
phenomena occur. Most notably, these are the confinement of color
charges and the spontaneous breaking of chiral symmetry (SBCS).
The former refers to the fact that quarks and gluons have never
been observed as individual particles, but only come in
"colorless" baryons, where three quarks carry an equal amount of
three different color charges, or mesons, where quark and
antiquark carry color charge and anticharge.

In the limit of vanishing bare quark masses, which is  quite good
approximation for the light up and down flavors, and, to a lesser
extent, for the heavier  strange flavor  ($m_u\approx5\,{\rm
MeV}$, $m_d\approx8\,{\rm MeV}$, and $m_s\approx125\,{\rm MeV}$),
the QCD Lagrangian is invariant under rotations of left-handed and
right-handed quarks in isospin (flavor) space ($SU(3)_L \times
SU(3)_R$ chiral symmetry of the QCD Lagrangian): \be
q_{L(R)}\longrightarrow
e^{i\sum_a\xi^a_{L(R)}\lambda_a}q_{L(R)}\quad
q_{L,R}=\frac{1}{2}(1\mp\gamma_5)q,\ee where $\lambda_a$ are the
Gell-Mann matrices being the generators of $SU(3)$ group. However,
the constituent quark mass (i.e., the mass of a bare quark dressed
by a virtual quark-gluon cloud) breaks the chiral symmetry,
because massive quarks can change their chirality so that it is no
longer conserved. A physical reason for the origin of the
constituent quark mass is that the QCD vacuum is not empty but
filled with various condensates of quark-antiquark and gluon
fields.  As a consequence, the light quarks acquire an effective
mass when propagate through the condensed vacuum,
  $m_q^\ast\simeq
G\langle 0|\bar qq|0\rangle\simeq 0.4\,\mathrm{GeV}$ ($G$ is the
effective quark coupling constant), being larger than the bare
quark masses by a factor of about 100. The theoretical
understanding of the mechanisms underlying confinement and SBCS
constitutes a major challenge in the contemporary particle and
nuclear physics research.

\section{QCD Phase Diagram }
\begin{figure}
\begin{center}
  \begin{tabular}{cc}
  \includegraphics[height=5.5cm]{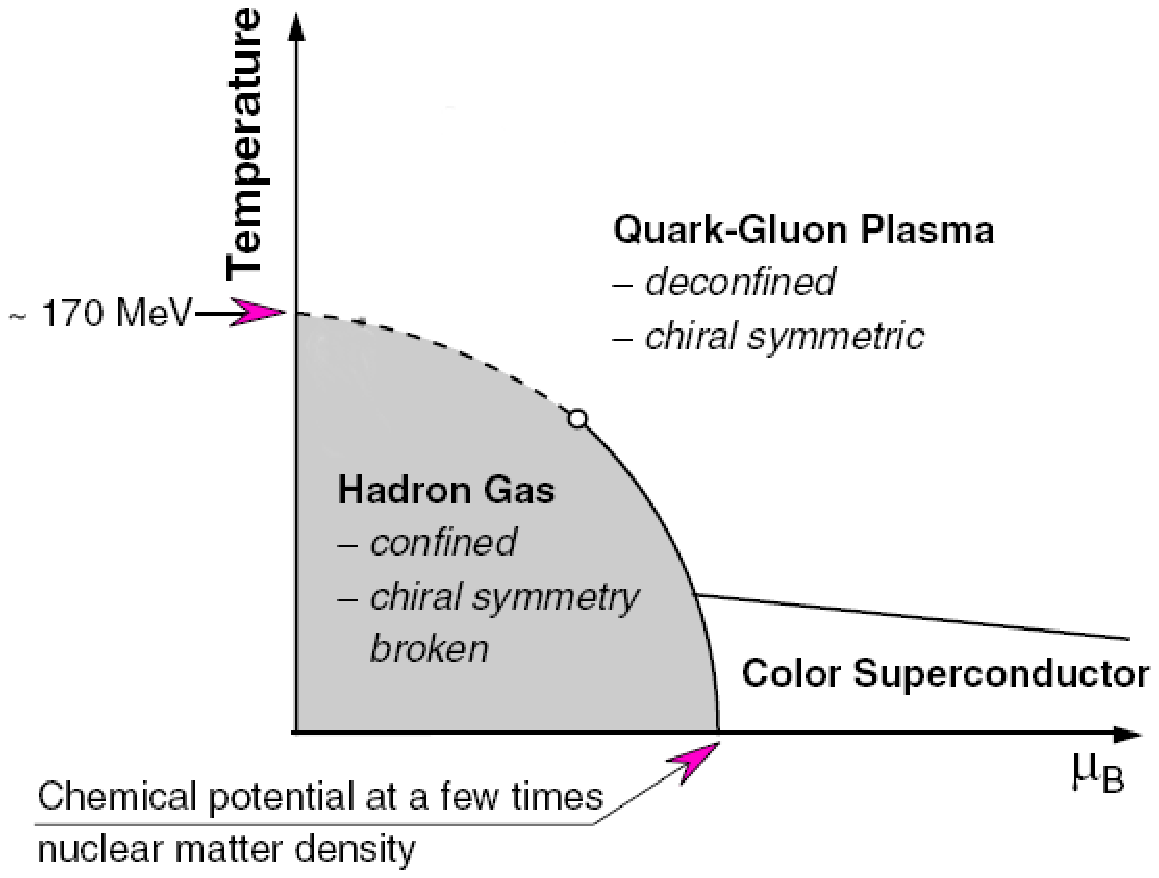}\quad & \quad\includegraphics[height=5.5cm]
  {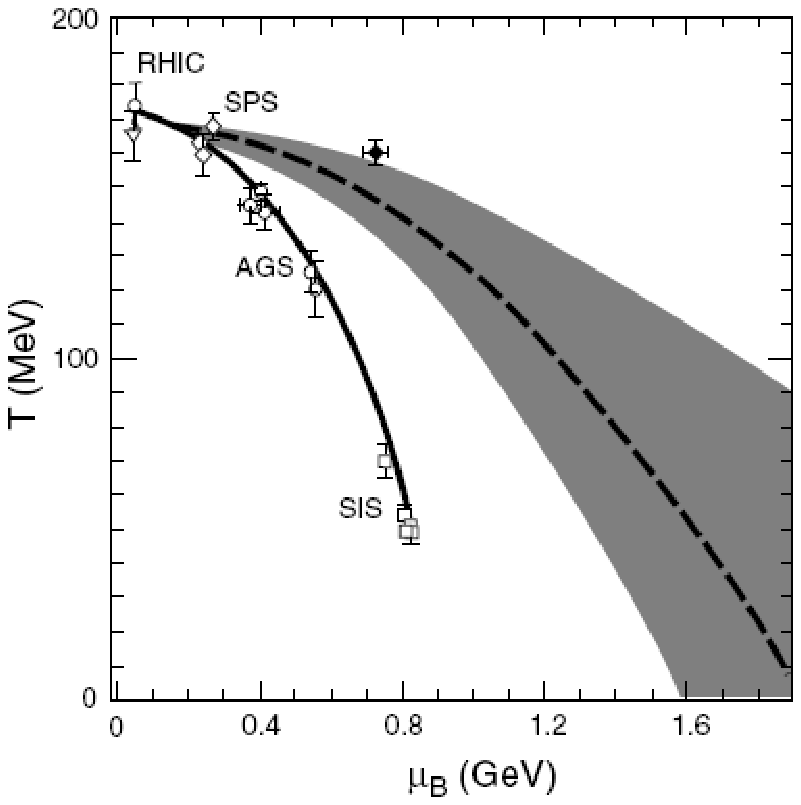}
  \end{tabular}
\end{center}
\vspace{-2ex} \caption{Left: Sketch of the QCD phase diagram.
Right:  LQCD results on the QCD-phase boundary (dashed line; the
grey band indicates the uncertainty), shown together with the
chemical freeze-out conditions for different experiments (open
symbols); the filled point represents the endpoint of the
crossover transition. } \label{fig1}\vspace{-2ex}
\end{figure}

Lattice QCD (LQCD)  provides a first-principles approach to
studies of large-distance, non-perturbative aspects of QCD. The
discrete space-time lattice that is introduced in this formulation
of QCD is a regularization scheme particularly well suited for
numerical calculations. LQCD  calculations  predict that at the
critical temperature $T_c\simeq170\,\mathrm{MeV}$, corresponding
to the energy density $\varepsilon_c \simeq
0.6\,\mathrm{GeV/fm^3}$, nuclear matter undergoes a phase
transition to QGP. In QGP, chiral symmetry is approximately
restored and quark masses are reduced from their large effective
values in hadronic matter to their small bare ones.  Since the
heavy quarks (charm, bottom, top) are too heavy to play any role
in the thermodynamics in the vicinity of the phase transition, the
properties of 2-flavour or 3-flavour QCD are of a most interest.
The order of the transition to  QGP as well as the value of the
critical temperature depend on the number of flavours and values
of the quark masses. One finds $T_c = (175 \pm 15)\,\mathrm{MeV}$
in the chiral limit of 2-flavour QCD and a $20\,\mathrm{MeV}$
smaller value for 3-flavour QCD. First studies of QCD with two
light quark flavours and a heavier (strange) quark flavour
indicate that the transition temperature for the physically
realized quark-mass spectrum is close to the 2-flavour value.
Although the transition is of the second order in the chiral limit
of 2-flavour QCD, and of the first order for 3-flavour QCD, it is
likely to be only a rapid crossover in the case of the physically
realized quark-mass spectrum~\cite{K}. The crossover, however,
takes place in a narrow temperature interval around $T_c\simeq
170\,\mathrm{MeV}$  which makes the transition between the
hadronic and plasma phases still well localized. The schematic
representation of the QCD phase diagram in the plane "baryonic
chemical potential  - temperature" is shown in Fig.~\ref{fig1}
(left)~\cite{JPG1}. The dashed line indicates a possible region of
the rapid crossover transition. The solid lines indicate likely
first-order transitions. The open circle gives the second-order
critical endpoint of the line of a first-order transition. At low
temperature and low density,  cold nuclear matter is a ground
state of the system and at larger density the high density phase
can be understood in terms of nearly degenerate, interacting Fermi
gases of quarks. The remnant attractive interaction between quarks
almost inevitably leads to quark-quark pairing and thus to the
formation of a colour superconducting phase with $\langle0|
qq|0\rangle\not=0$. All approximate model-based calculations
suggest that the transition between cold nuclear matter and a
color superconducting state is of the first order.

Fig.~\ref{fig1} (right) shows the results on determining the
position of the phase boundary obtained in LQCD. The dashed line
in this figure represents the extrapolation of the leading
$\mu_B^2$ order of the Taylor expansion of $T_c$  to larger values
of the chemical potential~\cite{All}. It is worth noting that,
within the statistical uncertainties, the energy density along
this line is constant and corresponds to
$\varepsilon_c\simeq0.6\,\mathrm{GeV/fm^3}$, i.e. to the same
value as that found from lattice calculations at $\mu_B = 0$. It
is thus conceivable that the QCD transition indeed sets in when
the energy density reaches a certain critical value.

\begin{figure}[tb] % fig 1
\begin{center}
\includegraphics[height=5cm,keepaspectratio]{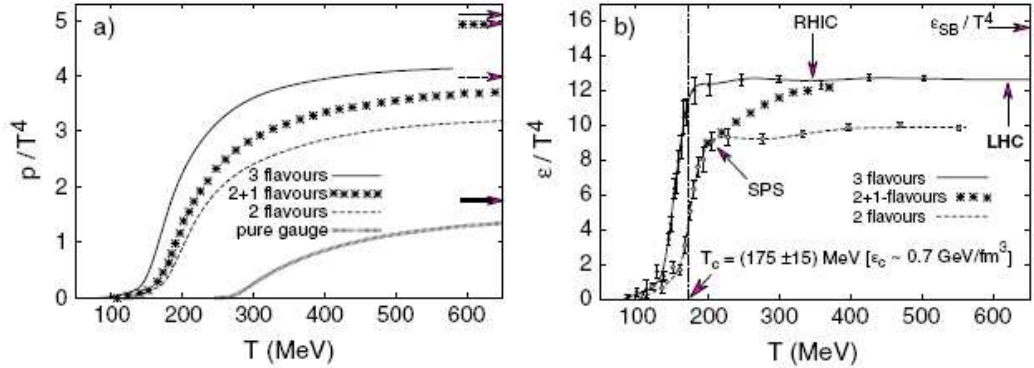}
\end{center}
\caption{The pressure (a) and energy density (b), in QCD with 0, 2
and 3 degenerate quark flavours as well as with two light (up and
down) flavours and a heavier (strange) quark flavour.  The arrows
on the right-side ordinates show the value of the Stefan-Boltzmann
limit for an ideal quark-gluon gas.} \label{fig2} %\vspace{-2ex}
\end{figure}

The pressure and energy density in QCD with 0, 2 and 3 degenerate
quark flavours as well as with two light (up and  down) flavours
and a heavier (strange) quark flavour are shown in
Fig.~\ref{fig2}. It is seen that near the crossover temperature
the energy density rapidly increases. The pressure, however, rises
more slowly near and above $T_c$, which indicates that significant
non-perturbative effects are to be expected at least up to
temperatures $T \simeq (2-3)T_c$. In fact, the observed deviations
from the Stefan-Boltzmann limit of an ideal quark-gluon gas are
quite significant even at temperatures  $T \simeq 5\,T_c$. As seen
from the right panel, in ultra-relativistic heavy ion collisions
(SPS, RHIC, LHC) one expects to attain energy densities which
reach and exceed the critical energy density $\varepsilon_c$, thus
making the QCD phase transition feasible  in the laboratory
conditions.

At the LHC, the critical energy density will be achieved by
colliding two lead nuclei at the center of mass energy $\sqrt s =
5.5\,\mbox{TeV}$ per nucleon pair. Different phases of heavy ion
collision can be described as follows~\cite{RH}. Incoming lead
nuclei at highly relativistic energies approach  each other,
subject to the substantial Lorentz contraction relative to their
transverse size. Upon initial impact of the nuclei, primordial
hard nucleon-nucleon collisions occur. Further reinteractions
induce the formation of QGP, whose pressure drives a collective
expansion and cooling, followed by hadronization and further
expansion in the hadronic phase. The particle abundances and hence
the chemical composition of the fireball are frozen almost
directly at hadronization, when inelastic processes cease
(chemical freeze-out).  At thermal freeze-out, the short-range
strong interactions cease  and the particle spectra are frozen.
Since the inelastic cross sections ($\sim 1\,\mathrm{mb}$) are
much smaller than the elastic ones ($\sim 100\,\mathrm{mb}$),
chemical freeze-out happens significantly before   thermal
freeze-out is reached.

Thus, the system created in heavy ion collision undergoes a fast
dynamical evolution from the extreme initial conditions to the
dilute final hadronic state. The objective is then  to identify
and to assess suitable QGP signatures, allowing to study the
properties of QGP.
\section{Probes of QGP}

The properties of the QGP state can be studied by means of a
variety of observables.  Mainly, we will be concentrated on
particles with the zero longitudinal momentum $p_z=0$ in the
center-of-mass frame of a nucleus-nucleus collision, the so-called
midrapidity ($y=0$) region, where one expects the largest energy
deposition of the interpenetrating nuclei. Thus, the main
kinematic variable is the transverse momentum $p_{\,T}$ of a
particle. In ALICE, the QGP observables (probes) are subdivided
into three classes: 1) soft probes with the typical momenta
$p_{\,T}\lesssim 2\,\mathrm{GeV/c}$, 2) heavy-flavour probes,
i.e., using the particles having c- and b-quarks, and 3)
high-$p_{\,T}$ probes in the momentum range about several tens
GeV/c. Further each of the classes will be illustrated  by a few
typical examples.

\subsection{Soft Probes of QGP}

\tit{Charged particle multiplicity.} One of the first measurements
in the ALICE physics programme will be the charged-particle
multiplicity per rapidity unit, or rapidity density, $dN_{ch}/dy$,
determined at midrapidity. The rapidity $y$ is defined according
to the equation $ y=\frac{1}{2}\ln\frac{E+p_z}{E-p_z},$ $E$ being
the energy and $p_z$ being the longitudinal momentum  on the
direction of the beam axis of a secondary particle in the
center-of-mass frame. The use of the rapidity as an independent
variable is useful since it changes by an additive constant  under
the longitudinal Lorentz boosts while the form of the rapidity
distribution remains unchanged. Alternatively, one can use the
pseudorapidity density $dN_{ch}/d\eta$,
$\eta=-\ln\left(\tan\frac{\theta}{2}\right)$ being the
pseudorapidity, related to the  angle $\theta$ between the
momentum of a secondary particle and beam axis. The ALICE detector
is optimized for the charged-particle density $dN_{ch}/dy=4000$
and its performance is checked in detailed simulations up to
$dN_{ch}/dy=8000$. Nowadays, the expected value at midrapidity for
LHC ranges between 1200 and 2900. The data on the charged particle
multiplicity allow one to get the estimate of the energy density
that could be reached  within the nuclei overlapping
zone\,\cite{B}. Such estimates show that the energy density at LHC
is $3-5$ times as many as the one $\varepsilon\approx 15\,
\mathrm{GeV/fm^3}$ reached at RHIC.

\tit{Hadron yields.} Another early measurement of ALICE will be
the identified hadron relative abundances. There are several
models for the description of hadroproduction. In particular, in
the thermal model~\cite{ABS}, the interacting system is considered
as a grand canonical ensemble of hadrons of various species formed
at a chemical freeze-out stage. The model parameters are
temperature and barionic chemical potential. This model is very
successful for the description of hadroproduction at RHIC
energies. Another theoretical model, the statistical hadronization
model~\cite{LR}, is based on the assumption that the expected
increase of strangeness in ultrarelativistic heavy ion collisions
could deviate from the grand canonical description. This deviation
is characterized by the strangeness phase space occupancy
$\gamma_s$ and at hadronization $\gamma_s>1$ (at the LHC, one
expects that $\gamma_s>5$). The data on the multiplicity
distribution and relative hadron abundances will allow one to
constrain the hadroproduction models.

\begin{figure}[tb] % fig 1
\begin{center}
\includegraphics[width=6cm,keepaspectratio,trim=0mm 0mm 0mm 0mm,
draft=false,clip]{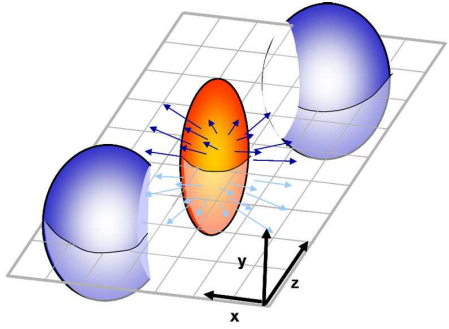}
\end{center}
\caption{The build-up of elliptic flow in a non-central heavy ion
collision.} \label{fig3}
\end{figure}

\tit{Particle momentum spectra. Elliptic flow.} In the
low-$p_{\,T}$ regime, the spectra of the most abundantly produced
hadrons ($\pi, K, p, \Lambda$, etc.) are well described by the
hydrodynamic models of  evolution  of an exploding
fireball~\cite{BO,CFBK}. In a noncentral collision of two heavy
nuclei, the initial nuclear overlap zone is spatially deformed
(see Fig.~\ref{fig3}, from Ref. [5]). Once the system thermalizes,
the pressure gradients along the short axis in the transverse
plane are larger than those along the long axis. As a result,
hydrodynamic expansion will be stronger along the $x$-axis
relative to the $y$-axis that leads to the  build-up of elliptic
flow in the collective matter expansion. The corresponding
azimuthal asymmetry in the  hadron transverse momentum spectra is
characterized by the elliptic flow coefficient $v_2$, defined by
the expansion
 \be \frac{d^3N_h}{d^{2}p_{\,T}dy}=\frac{d^2N_h}{\pi
dp_{\,T}^2dy}\Bigl(1+2v_2(p_{\,T})\cos2\varphi+\dots\Bigr) \ee (at
midrapidity, the system is mirror symmetric in the transverse
$x$-$y$ plane and the odd Fourier harmonics in the azimuthal angle
$\varphi$ are missing). If the matter produced in the reaction
zone rescatters efficiently, the spatial anisotropy of the
pressure gradients will be transferred into the anisotropy in the
momentum distribution.  However, if rescattering among the
produced particles is weak, an initial period of almost free
streaming will reduce the spatial anisotropy and thus reduce the
system's ability to convert the spatial anisotropy into the
momentum anisotropy.  Therefore, a large positive $v_2$ can only
be obtained if the thermalization of the medium is rapid enough.
In this way, the elliptic flow coefficient $v_2(p_{\,T})$ serves,
in principle, as a quantitative tool for  measuring the
thermalization time $\tau_0$. According to the RHIC data,  the
experimentally measured $v_2(p_{\,T})$ for various hadrons ($\pi,
K, p, \Lambda$) is best described within the hydrodynamic approach
when implementing a thermalization time of
$\tau_0=0.5-1\,\mathrm{fm/c}$. Besides, for hydrodynamics to build
up the observed $v_2$, the viscosity of the formed medium must
remain very small.

\subsection{Heavy-flavor probes of QGP}

\tit{Heavy quarkonia production.} Heavy quarkonia of interest
include charmonia and bottomonia, where charmonia (bottomonia) are
the bound states of the charm (bottom) quark $c$ ($b$) and
antiquark $\bar c$ ($\bar b$),  corresponding to different quantum
numbers of a bound state. For specifying the quantum numbers, we
use the usual spectroscopic notations, i.e., $^{2S+1}L_J$ with $S$
being the total spin, $L$ total orbital momentum, $J$ total
angular momentum of a quark-antiquark pair. Some examples of
charmonia states:  $J/\psi$ meson (mass $M=3.097\, \mathrm{GeV}$)
- $c\bar c$ in the $^3S_1$ state, $\psi'$ meson ($M=3.685\,
\mathrm{GeV}$) - first radial excitation above the lowest lying
$^3S_1$ state of $c\bar c$, $\eta_c$ meson ($M=2.98\,
\mathrm{GeV}$) - $c\bar c$ in the $^1S_0$ state, $\chi_{cJ}$ meson
($M=3.41, 3.51, 3.56\, \mathrm{GeV}$ at $J=0,1,2$, respectively) -
$c\bar c$ in the $^3P_J$ state. Examples of bottomonia states:
$\Upsilon$ meson ($M=9.46 \, \mathrm{GeV}$)~- $b\bar b$  in the
$^3S_1$ state, $\Upsilon',\Upsilon'',\Upsilon'''$ (with the masses
10.02, 10.40 and 10.55 GeV, respectively) - three first radial
excitations above  the ground $^3S_1$ state of $b\bar b$.

\begin{figure}
\begin{center}
  \begin{tabular}{cc}
  \includegraphics[width=6.8cm]{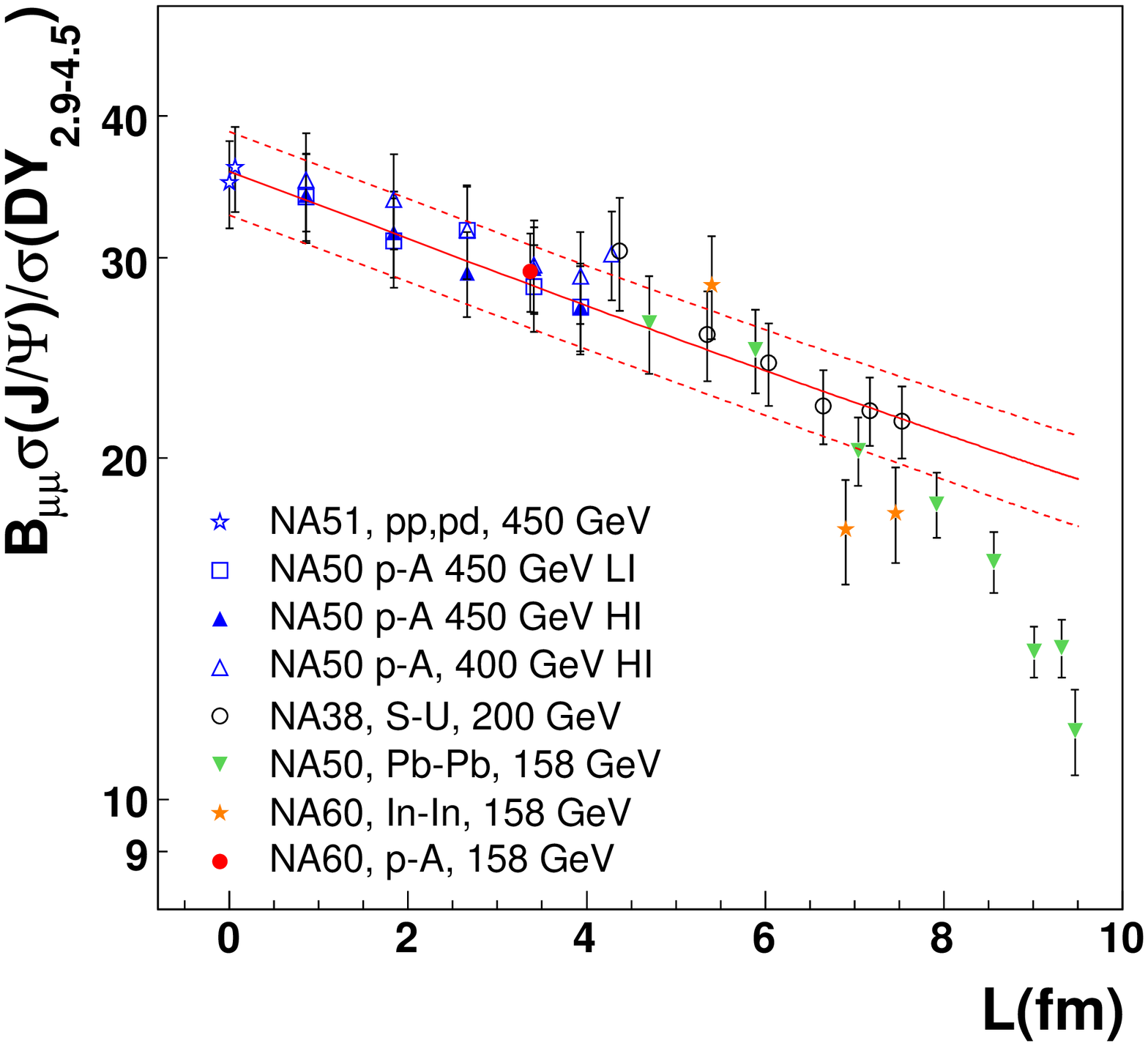} & \includegraphics[width=6.8cm]{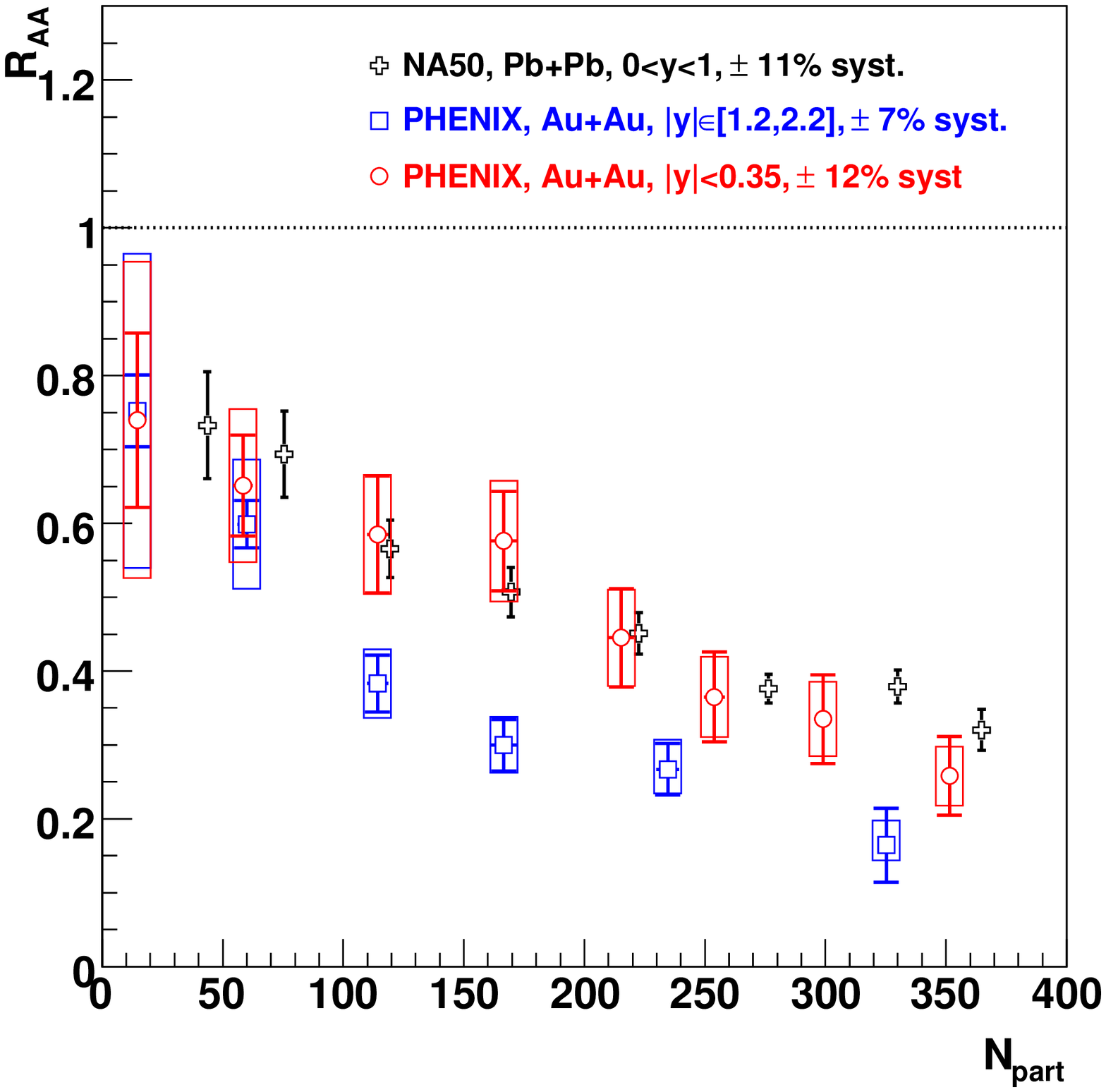}
  \end{tabular}
\end{center}
%\vspace{-2ex}
\caption{Left: $J/\psi$ yields normalized by
Drell-Yan process, as a function of the nuclear thickness $L$, as
measured at the SPS. Right: $J/\psi$ nuclear modification factor
for the most energetic  SPS (Pb-Pb) and RHIC (Au-Au) collisions,
as a function of the number of participants $N_{part}$. }
\label{fig4}%\vspace{-2ex}
\end{figure}

In nucleus-nucleus collisions, quarkonium suppression is expected
to occur due to the Debye screening of the heavy quark interaction
in  QGP. For example, increase of the energy density reached in
the collisions leads to break up of first  $\psi'$ and $\chi_c$,
and then   $J/\psi$ mesons. Further we consider in more detail the
suppression of $J/\psi$ mesons, being one of the key probes of the
QGP formation in heavy ion collisions\,\cite{MS}. Soon it was
realized that besides melting $J/\psi$ mesons in QGP due to the
screening of the color charge, there are also a few competing
mechanisms which could explain the suppression of $J/\psi$
production in heavy ion collisions. These mechanisms are referred
to as cold nuclear matter effects (CNM) and  include: 1)
absorption of $J/\psi$ by nuclear fragments from colliding nuclei;
2) shadowing of low momentum partons (the depletion of low
momentum partons in nucleons bound in nuclei as compared to free
nucleons). The last point is important because the gluon fusion is
one of the main production mechanisms of $J/\psi$s  and the
$J/\psi$ yield is therefore sensitive to gluon shadowing. The
measurements of the $J/\psi$ yield at CERN SPS at
$\sqrt{s}=17.3\,\mathrm{GeV}$, depicted in the left panel of
Fig.~\ref{fig4}, show the anomalous suppression beyond the CNM
effects in the central collisions when the number of participating
nucleons is the largest\,\cite{S}. This can be considered as a
signature of melting of the charmonium states in QGP.

However, measurements of the $J/\psi$ suppression by PHENIX
collaboration at RHIC lead to some surprising features. The
$J/\psi$ suppression can be characterized by the nuclear
modification factor
$$R_{AA}(p_{\,T},y)=\frac{d^2N_{J/\psi}^{AA}/dp_{\,T}dy}{N_{coll}d^2N_{J/\psi}^{pp}/dp_{\,T}dy},$$
 obtained by normalizing the $J/\psi $ yields in heavy ion
collisions by the $J/\psi$ yields in $p+p$ collisions at the same
energy times the average number of binary inelastic
nucleon-nucleon collisions. This ratio characterizes the impact of
the medium on the particle spectrum. If heavy ion collision is a
superposition of independent $N_{coll}$ inelastic nucleon-nucleon
collisions, then $R_{AA}=1$, whereas $R_{AA}<1$ ($R_{AA}>1$)
corresponds to the case of the $J/\psi$ suppression (enhancement).
Fig.~\ref{fig4}~(right) shows the $p_T$ integrated nuclear
modification factor for CERN SPS and RHIC PHENIX
experiments~\cite{Al,A}. There are two surprising results in these
measurements. First, the midrapidity suppression (the red boxes)
in PHENIX is lower than the forward rapidity suppression (blue
boxes) despite the experimental evidence that the energy density
is higher at midrapidity than at forward rapidity, and, hence, one
could expect that at midrapidity the $J/\psi$s should be more
suppressed. Secondly, the nuclear modification factor $R_{AA}$ at
midrapidity in PHENIX (red boxes) and SPS (black crosses) are in
agreement within error bars although
 the energy density reached at RHIC is larger than the one reached at SPS.

 Possible explanations of these features could be the
following~\cite{RGC}. 1. Regeneration of $J/\psi$s in the hot
partonic phase from initially uncorrelated $c$ and  $\bar c$
quarks  (quark coalescence model).  The underlying idea is that at
midrapidity there are more $c$ and $\bar c$ quarks to regenerate
than at forward rapidity that could explain less suppression of
$J/\psi$s at midrapidity.  Note that the total number of initial
$c\bar c$ pairs is larger than 10 in the most central Au+Au
collisions at RHIC. 2. $J/\psi$ production could be more
suppressed at forward rapidity due to the nuclear shadowing
effects. Standard gluon shadowing parametrizations do not tend to
produce such an effect but they are poorly constrained by the data
and further saturation effects are not excluded.

Thus,  the ideal signature of deconfinement would be an onset of
the $J/\psi$ suppression. However, such an onset need to be looked
at after the CNM effects are properly taken into account. Besides,
the role of quark coalescence should be clarified at energies
higher than the SPS energy. This is especially important at the
LHC because the number of $c\bar c$ pairs produced in one central
Pb-Pb collision is expected to be larger than 100. One of
possible experiments aimed to verify  the quark coalescence model
is the measurement of the $J/\psi$ elliptic flow. The idea is that
if charmonia were produced by coalescence of charm quarks, they
should inherit somehow  their flow, resulting in a higher $v_2$
than in the case of the direct production in hard
collisions~\cite{CS}.

It is also worth to note that some level of the $J/\psi$
suppression could originate with the reduced feed-down to $J/\psi$
from excited charmonium states ($\psi',\chi_c$), which melt just
above the QGP transition temperature, and from $B$ mesons. Thus,
further studies are necessary to make the definite conclusions
about the $J/\psi$ production pattern.

Bottomonia are also of a special  interest for the studies at the
ALICE. One could suppose that bottomonia might be easier to
understand than charmonia. Since only about 5 $b\bar b$ pairs are
expected to be produced in a single central Pb-Pb collision,
regeneration should play much less role in the beauty sector at
the LHC. Besides, having higher masses, bottomonia originate from
higher momentum partons and will less suffer from shadowing
effects.  These features should ease the  separation of the
anomalous suppression in the $\Upsilon$'s family.

\begin{figure}[tb]\begin{center}
\includegraphics[height=5.5cm,keepaspectratio]{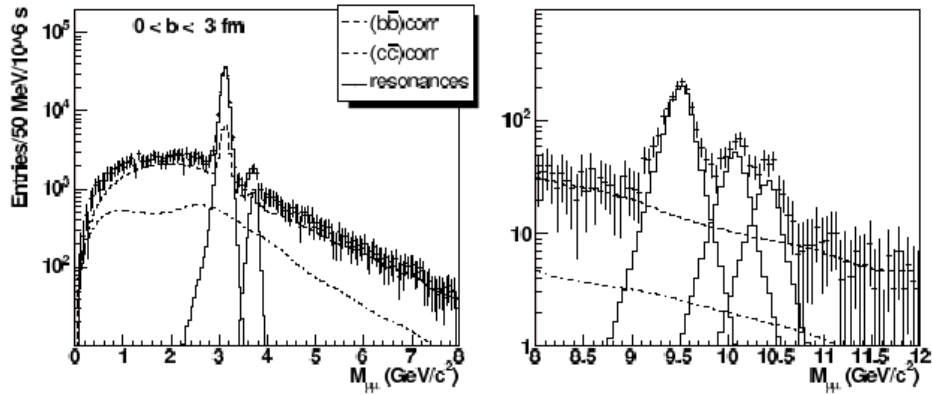}
\end{center}
\caption{Left: Dimuon invariant mass spectra (after subtraction of
the uncorrelated background) in central Pb-Pb collisions for one
year of data taking ($10^6$ s) and  luminosity $5\cdot
10^{26}\,\mathrm{cm}^{-2}\mathrm{s}^{-1}$ in the  $J/\psi$ mass
region. Right:~Same as  on the left, but in the $\Upsilon$ mass
region.
 } \label{fig5}
\end{figure}

In ALICE, quarkonia will be detected both in the dielectron
channel at midrapidity and in the dimuon channel at forward
rapidity. For example, in the dimuon channel, signals of about
$7\cdot 10^5$ $J/\psi$s and $10^4$ $\Upsilon$s are expected in the
most central Pb-Pb collisions for one  year of data taking at
nominal luminosity (see Fig.~\ref{fig5} for the corresponding
dimuon invariant mass spectra~\cite{JPG2}).

\tit{Nuclear modification factor for $D$ and $B$ mesons.} The
heavy-light mesons, $D =(c\bar q)$,   and $B =(b\bar q)$, are
composed of a heavy quark $Q=(c, b)$ and a light antiquark $\bar
q=(\bar u,\bar d)$.     When traversing the dense matter created
in nucleus-nucleus collisions, the initially-produced hard partons
lose energy mainly on account of medium-induced gluon radiation.
Quarks are predicted to lose less energy than gluons which have a
higher color charge. The heavy quarks at intermediate $p_{\,T}$
will lose less energy as compared with the light quarks at the
same momentum, since bremsstrahlung off  accelerated heavy quarks
is suppressed by a large power of their mass relative to light
quarks, $\sim (m_q/m_Q)^4$. Some of heavy quarks will be
fragmented into $D$- and $B$-mesons. The influence of the medium
on their momentum spectrum is characterized by the nuclear
modification factor $R_{AA}(p_{\,T})$. Due to the above features,
one should observe a pattern of gradually decreasing $R_{AA}$ when
going from the mostly gluon-originated light-flavor hadrons
($h^\pm$ and $\pi^0$) to $D$ and then to $B$ mesons~\cite{Ar}:
$R_{AA}^h\lesssim R_{AA}^D\lesssim R_{AA}^B$. The enhancement
above the unity of the heavy-to-light ratio
$R_{AA}^{D/h}=R_{AA}^D/R_{AA}^h$ probes the color charge
dependence of the parton energy loss while the ratio
$R_{AA}^{B/D}=R_{AA}^B/R_{AA}^D$ probes the mass dependence of the
parton energy loss.

\subsection{High-$p_{\,T}$ probes of QGP}

\tit{Jet production.} In the high-$p_{\,T}$ regime, the mechanism
of hadroproduction changes (as compared to the production of
hadrons from recombination of partons at $p_{\,T}$ up to a few
GeV/c). The high-$p_{\,T}$
 partons produced in the initial stage of a
nucleus-nucleus collision undergo multiple scatterings inside the
collision region and then fragment into a jet, representing a
collimated spray of hadrons. These multiple scatterings are
expected to induce modifications in the properties of the produced
jet, which probe the state of hot and dense matter formed in the
initial stage of  heavy ion collision. This is the main motivation
for studying jet production in nucleus-nucleus collisions.
Back-to-back jets are routinely observed in high-energy collisions
of elementary particles, but are difficult for identification in
the high-multiplicity environment of  heavy ion collision.
However, a jet typically contains a leading particle which carries
most of the momentum of the parent parton. High-$p_{\,T}$ spectra
in heavy ion collisions thus essentially determine the
modification underlying the production of the leading hadron in a
jet. This modification can be characterized by the nuclear
modification factor $R_{AA}^h(p_{\,T})$ for the leading hadron.
RHIC data  obtained in central Au+Au collisions show that
high-$p_{\,T}$ spectra of hadrons are strongly
suppressed~\cite{R}, e.g., for pions, $R_{AA}^\pi\sim0.2$ up to
$p_{\,T}\sim20\,\mathrm{GeV/c}$. These observations provide strong
evidence for jet quenching: high-$p_{\,T}$ partons lose energy
while traversing quark-gluon plasma  due to induced gluon
radiation and elastic scattering. Note also that at extreme
quenching scenarios a full jet reconstruction should provide a
better sensitivity to the medium properties. The modification of
the jet structure is then expected to be seen in the decrease of
the number of particles carrying a high fraction of the jet
energy, the increase of the number of the low energy particles and
broadening of the distribution of  the jet-particle momenta
perpendicular to the jet axis.

\begin{figure}[tb]\begin{center}
\includegraphics[width=8cm,keepaspectratio]{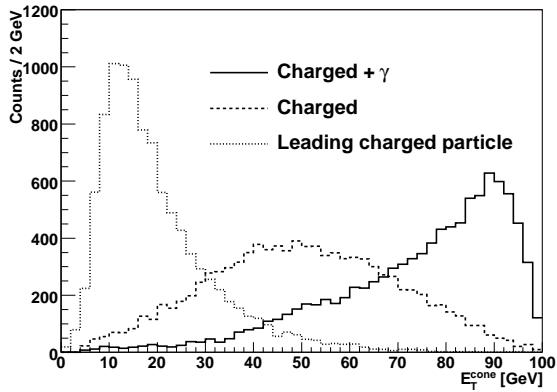}
\end{center}
\caption{Distribution of reconstructed energy for simulated 100
GeV jets. The results for a charged leading particle, all charged
paticles, and all charged particles and photons included in the
analysis are shown. } \label{fig6}
\end{figure}

At the LHC, one anticipates high statistics of jets with the
energies at which they can be distinguished from the underlying
event. For example, the rate of 100 GeV jets is predicted to be
around $10^6$ per month in central Pb-Pb collisions.
Fig.~\ref{fig6} shows the spectra of the reconstructed jet energy
for jets of 100 GeV, using a leading charged particle only, all
charged particles, and charged particles plus photons. It
illustrates an improvement in the jet energy reconstruction after
upgrading ALICE with  the Electro-Magnetic Calorimeter
(EMCAL)~\cite{JPG2}.

\tit{Direct photons.} Direct photons are defined as photons that
do not originate from the decay of other particles. Owing to their
small electromagnetic coupling, photons once produced do not
interact with the surrounding matter and thus probe the state of
matter at the time of their production.  Early in the collision,
the so-called "prompt" photons are produced by hard parton-parton
scattering in the primary nucleus-nucleus collisions. They
dominate the photon spectrum at $p_{\,T}\gtrsim10\,\rm{GeV/c}$.
Although their rate decreases as an inverse power of $p_{\,T}$,
photons up to several hundred GeV are expected to be detected at
LHC.  The production cross sections of prompt photons are
calculated in perturbative QCD from the basic processes  by a
convolution with the parton density distribution in the nucleus.
The basic processes include  the direct processes such as Compton
scattering ($g+q\, (\bar q)\rightarrow \gamma+q\, (\bar q))$ or
annihilation ($q\bar q\rightarrow \gamma g$), and the
bremsstrahlung process in which the photon is produced in the
fragmentation of a quark $q\rightarrow\gamma$, or a gluon
$g\rightarrow\gamma$.  Medium effects in A-A collisions modify the
production cross sections of prompt photons as measured in $pp$
collisions: nuclear shadowing and in-medium parton energy loss
lead to a suppression of the yield, whereas the intrinsic
transverse momentum distribution of the partons  and
medium-induced photon radiation from quark jets  enhance the
yield. An important background to prompt photon production is the
decay $\pi_0\rightarrow\gamma\gamma$, produced in similar hard
partonic processes.

A promising method to probe the medium effects in heavy ion
collisions is to use the  jets, originating from Compton or
annihilation process. Since prompt photons are produced in parton
collisions,  in which at the central rapidity region the final
state photon and parton are emitted almost back-to-back, such jets
can be tagged with prompt photons, emitted oppositely to the jet
direction. Medium effects then can be identified through the
determination of the nuclear modification factor $R_{FF}$ related
to the ratio of the fragmentation functions measured in A-A and
$pp$ collisions and scaled with the number of binary
nucleon-nucleon collisions. The fragmentation function of a
photon-tagged jet is the distribution of charged hadrons within
the jet as a function of $p_{\,T}/E_\gamma$ ($E_\gamma$ is the
photon energy). Medium effects in heavy ion collisions can be also
probed by studying photon-hadron and photon-photon correlation
functions.

In conclusion, the ALICE experiment at the LHC is aimed to study
the physics of strongly interacting matter at extreme energy
densities and temperatures, where a new state of matter,
quark-gluon plasma, is expected to be reached.  Due to its
excellent track, vertex-finding and particle-identification
capabilities, ALICE will be able to study the properties of QGP by
means of a whole set of different and complementary observables.

\end{document}